\documentclass[12pt]{article}
\usepackage{latexsym}
\usepackage{amsfonts}
\usepackage{graphicx}

\def\be{\begin{equation}}
\def\ee{\end{equation}}
\def\beq{\begin{eqnarray}}
\def\eeq{\end{eqnarray}}

\newsavebox{\uuunit}
\sbox{\uuunit}
    {\setlength{\unitlength}{0.825em}
     \begin{picture}(0.6,0.7)
        \thinlines
        \put(0,0){\line(1,0){0.5}}
        \put(0.15,0){\line(0,1){0.7}}
        \put(0.35,0){\line(0,1){0.8}}
       \multiput(0.3,0.8)(-0.04,-0.02){12}{\rule{0.5pt}{0.5pt}}
     \end {picture}}

\begin{document}

\begin{titlepage}
\begin{flushright}
ULB-TH/03-25\\

\end{flushright}
\vskip 1.0cm

\begin{centering}

{\large {\bf Einstein billiards and spatially homogeneous cosmological
models}}

\vspace{.5cm}

Sophie de Buyl\footnote{Aspirant du Fonds National de la Recherche
Scientifique,
Belgique}, Ga\"{\i}a Pinardi and Christiane Schomblond
\\
\vspace{.7cm} {\small Physique Th\'eorique et Math\'ematique,\\
Universit\'e Libre
de Bruxelles, \\ C.P. 231, B-1050, Bruxelles, Belgium.\\sdebuyl@ulb.ac.be,
gpinardi@ulb.ac.be, cschomb@ulb.ac.be}

\vspace{.5cm}

\end{centering}

\begin{abstract}
In this paper, we analyse the Einstein and Einstein-Maxwell billiards
for all spatially homogeneous cosmological models corresponding to 3 and 4
dimensional real unimodular Lie algebras and provide the list of those
models which are chaotic in the Belinskii, Khalatnikov and Lifschitz
(BKL) limit. Through the billiard picture, we confirm that, in $D=5$
spacetime dimensions, chaos is present if off-diagonal metric elements
are kept: the finite volume billiards can be
identified with the fundamental Weyl chambers of hyperbolic Kac-Moody algebras.
The most generic cases bring in the same algebras as in the inhomogeneous
case, but other algebras appear through special initial conditions.
\end{abstract}

\vfill
\end{titlepage}
\section{Introduction}
\setcounter{equation}{0} \setcounter{theorem}{0}
\setcounter{lemma}{0}

It has recently been shown \cite{DH,DHJN} that the classical dynamics of the
spatial scale factors (and of the dilatons if any) of
$D$-dimensional gravity coupled to $p$-forms can be described, in the
vicinity of a spacelike singularity, as a billiard motion, i.e. as the
relativistic motion of a ball inside a region of hyperbolic space
bounded by hyperplanes on which it undergoes elastic bounces. These
results generalize the work by Belinskii, Khalatnikov and Lifshitz (BKL)
\cite{BKL} that discovered the chaotic behaviour of the generic solution
of the vacuum Einstein equation, in four dimensional spacetime, near such
a singularity. These authors showed that, near a spacelike singularity, a
decoupling of the spatial points occurs which hugely simplifies the
dynamical equations of the spatial metric, in the sense that,
asymptotically, they cease to be partial differential equations to simply
become, at each spatial point, a set of second-order non linear ordinary
differential equations with respect to time.

These equations coincide with the dynamical equations of some spatially
homogeneous cosmological models which possess some of the qualitative
properties of more generic solutions. For
$D=4$, the spatially homogeneous vacuum models that share the chaotic
behaviour of the more general inhomogeneous solutions are labelled as
Bianchi type IX and VIII; their homogeneity groups are respectively
$SU(2)$ and
$SL(2,\mathbb{R})$. In higher spacetime dimensions, i.e. for $5\leq D\leq
10$, one also knows chaotic spatially homogeneous cosmological models but
none of
them is diagonal \cite{B}. In fact, diagonal models are too restrictive to be
able to reproduce the general oscillatory behaviour but, as shown e.g. in
\cite{DHdR}, chaos is restored when non-diagonal metric elements are taken into
account.

A systematic way to study the asymptotic behaviour of solutions of
Einstein's equations in the neighborhood of a spacelike singularity and
in any spacetime dimension is provided by the Hamiltonian approach
developed in
\cite{DHN} in which gravity is coupled to a collection of $p$-forms. In
this framework, a billiard description of the
asymptotic evolution of the scale factors is naturally set up and
chaos follows from the fact that the billiard's volume is finite. In
many interesting examples, some of them related to supergravity
models
\cite{DHN, DdBHS}, the billiard can be identified with the fundamental
Weyl chamber of a Kac-Moody algebra and the reflections against the
billiard walls with the fundamental Weyl reflections which generate the
Weyl group (or Coxeter group): accordingly, the finiteness of the volume,
hence also chaos, rely upon the hyperbolic character of the
underlying Kac-Moody algebra; this last property is established through
its Cartan matrix or equivalently in its Dynkin diagram.

The purpose of this paper is to analyse the billiard evolution of
spatially homogeneous non-diagonal cosmological models in $D=4$ or
$5$ spacetime dimensions, in the Hamiltonian formalism mentionned
hereabove. Since one knows that the full field content of the
theory is important in the characterization of the billiard, we
compare the pure Einstein gravity construction to that of the
coupled Einstein-Maxwell system. The gravitational models we are
interested in are in a one-to-one correspondence with the real Lie
algebras - a complete classification based on their structure
constants exists for $d=3$ and $d=4$, $(D=d+1)$\cite{LL,
McC}\footnote{We will use the notations of MacCallum, we refer to
\cite{McC} for translation to other notations} - and we restrict
our analysis to the unimodular ones, because only for such models
can the symmetries of the metric be prescribed at the level of the
action. We proceed along the lines defined in \cite{DHN} but with
a special concern about the r\^ole played by the constraints.
Indeed, while in the general inhomogeneous case, the constraints
essentially assign limitations on the spatial gradients of the
fields without having an influence on the generic form of the BKL
Hamiltonian, in the present situation, they precisely relate the
coefficients that control the walls in the potential.
Consequently, the question arises whether they can enforce the
disappearing of some (symmetry or electric or magnetic) walls as
already do the vanishing structure constants with gravitational
walls. They could prevent the generic oscillatory behaviour of the
scale factors. The answer evidently depends on the Lie algebra
considered and on its dimension: for example, while going from the
Bianchi IX model in $d=3$ to the corresponding $U3S3$ model in
$d=4$, the structure constants remain the same but the momentum
constraints get less restrictive. Hence generic behaviour is
easier to reach when more variables enter the relations. We find
that, except for the Bianchi IX and VIII cases in $D=4$, symmetry
walls (hence off-diagonal elements) are needed to close the
billiard table: thereby confirming, in the billiard picture,
previous results about chaos restoration. Moreover, we find that
when the billiard has a finite volume in hyperbolic space, it can
again be identified with the fundamental Weyl chamber of one of
the hyperbolic Kac-Moody algebras. In the most generic situation,
these algebras coincide with those already relevant in the general
inhomogeneous case. However, in special cases, new rank 3 or 4
simply laced algebras are exhibited.

The paper is organized as follows. We first adapt to the spatially
homogeneous case that part of the general Hamiltonian formalism set up in
\cite{DHN} necessary to understand how, at the BKL limit, the billiard
walls arise in the potential. We explicitly write down the form of the
momentum constraints in the generalized Iwasawa variables and
analyse their meaning for each of the 3 and 4 real unimodular Lie
algebras as well as their impact on the billiard's shape.
For the finite volume billiards, we compute the scalar products
of the gradients of the dominant walls using the metric defined by the kinetic
energy and show that the matrix
 \be A_{AB} =
2 \frac{(w_A\vert w_B)}{(w_A\vert w_A)}\quad\mbox{where}(w_A\vert w_B) =
G^{ab}\,w_{Aa} w_{Bb}\ee
 is the generalized Cartan matrix of an hyperbolic
Kac-Moody algebra.

\section{General setting}
\setcounter{equation}{0} \setcounter{theorem}{0}
\setcounter{lemma}{0}

\subsection{Spatially homogeneous models, Hamiltonian}

In this paper, we are specially interested in $d=3$ and
$d=4$ dimensional spatially homogeneous models equipped with a
homogeneity group simply transitively acting; these models are known to be in a
one-to-one correspondence with the $3$ and
$4$ dimensional real Lie algebras and have been completely
classified\cite{McC,RS}. We restrict our analysis to the
unimodular algebras, i.e. those whose adjoint representation is
traceless\footnote{The group Adjoint representation is unimodular.}, that is
$C^i_{\,\,ik}=0$ , since only for these homogeneous models do the
equations of motion follow from a reduced Hamiltonian action in which the
symmetry of the metric is enforced before taking
variationnal derivatives.

We work in a pseudo-Gaussian gauge
defined by vanishing shift $N^i = 0$ and assume the $D=d+1$ dimensional
spacetime metric
of the form
\be ds^2 = - (N
dx^0)^2 + g_{ij}(x^0)\,
\omega^i\,\omega^j\ee where $x^0$ is the time coordinate, $t$ is
the proper time, $dt = -N \,dx^0$, and $N$ is the
lapse. For definiteness and in agreement with the choice made in \cite{DHN},
we will assume that the spatial singularity occurs in the past, for $t=0$.

The gravitational dynamical variables
$g_{ij}$ are the components of the $d$ dimensional spatial metric in the
time-independent co-frame
$\{\omega^i =
\omega^i_{\,\,j}\,dx^j\}$ invariant
under the group transformations
\be d\omega^i = -\frac{1}{2}\,C^i_{\,\,\,jk}\,\omega^j\wedge \omega^k;\ee
the $C^i_{\,\,\,jk}$'s are the group structure constants. The metric
$g_{ij}(x^0)$ depends only on time and may contain
off-diagonal elements. With use of $g \equiv det\,g_{ij}$, one defines
the rescaled lapse as $\tilde N = N/\sqrt{g}$.

In the spatially homogeneous Einstein-Maxwell system, there is besides the
metric, an electromagnetic
$1$-form potential $A$ and its $2$-form field strength $F=dA$. In
the temporal gauge $A_0=0$, the potential reduces to
\be A =  A_j\,\omega^j.\ee In the Hamiltonian framework, we assume the
potential itself to be spatially homogeneous\footnote{This is more restrictive
than requiring spatial homogeneity of the field strength; in the
present analysis the difference only arises with regard to the magnetic walls
which are always subdominant.} so that its space components in the
$\omega^j$ frame are functions of
$x^0$ only: $A_j = A_j (x^0)$. Accordingly, its field strength takes the
special form
\be F= dA = \partial_0\, A_j\,dx^0\wedge \omega^j -
\frac{1}{2}\,A_i\,C^i_{\,\,\,jk}\,\omega^j\wedge \omega^k\ee which shows the
links between the components of the magnetic field and the structure
constants. Hence, from the Jacobi identity, one infers that \be
F_{i[j}\,C^i_{\,\,\,k\ell]} = 0.\label{cmag}\ee

The first order action for the homogeneous Einstein-Maxwell
system can be obtained from the $D$ dimensional Hilbert-Einstein action in
ADM form after space integration has been carried out; this operation
brings in a constant space volume factor that will be ignored hereafter.
The action is given by
\be
S[g_{ij},\pi^{ij},A_j,\pi^j] = \int dx^0 \big( \pi^{ij}{\dot g}_{ij} +
\pi^j{\dot A}_j - {\tilde N} H
\big).\label{action}\ee The Hamiltonian ${\tilde N}H$ reads as
\beq
{ H} &=& {K} + { M} \\ {K} &=&
\pi^{ij}\pi_{ij}-\frac{1}{d-1}\pi^i_{\,\,i}\pi^j_{\,\,j}+\frac{1}{2}\pi^j
\pi_j\\ { M} &=&-gR +
\frac{1}{4} F_{ij}F^{ij}\eeq
where $R$ is the spatial curvature scalar defined, in the unimodular
cases, by the following combination of structure constants and metric
coefficients
\be R = -\frac{1}{2}\,( C^{ijk}\,C_{jik} +
\frac{1}{2}\,C^{ijk}\,C_{ijk}\,) ,\ee where
\be C_{ijk} = g_{i\ell}\,C^\ell_{\,\,\,jk}\quad\mbox{and}\quad C^{ijk} =
g^{j\ell}\,g^{km}\,C^i_{\,\,\,\ell m}.\label{struc2}\ee

The equations of motion
are obtained by varying the action (\ref{action}) with respect to the
spatial metric components $g_{ij}$, the spatial $1$-form components $A_j$
and their respective conjugate momenta $\pi^{ij}$ and $\pi^j$. The
dynamical variables still obey the following constraints:
\beq {H}&\approx & 0\quad \mbox{(Hamiltonian constraint)}\\ {H}_i&=&
 - C^j_{\,\,\,ik}\,\pi^{k}_{\,\,j}+
\pi^j F_{ij}\approx 0\quad \mbox{(momentum constraints)};\label{momentum}
\eeq
notice that the Gauss law for the electric
field is identically satisfied on account of the unimodularity condition.

\subsection{Generalized Iwasawa variables}

In order to develop the billiard analysis, it is necessary to
change the variables, i.e. to replace the metric components $g_{ij}$ by
the new variables
$\beta^a$ and ${\cal N}^a_{\,\,i}$, defined through the Iwasawa
matrix decomposition \cite{DHN}
\begin{equation}  g = {\cal N}^T {\cal A}^2{\cal
N}\label{Iwas}\end{equation}
 where ${\cal N}$ is an upper triangular matrix with
$1$'s on the diagonal and ${\cal A}$ is a diagonal matrix with positive
entries parametrized as
\begin{equation} {\cal A} = exp (-\beta),
\qquad
\beta = diag (\beta^1, \beta^2,..., \beta^d).\end{equation} The explicit
form of (\ref{Iwas}) reads \be g_{ij} = \sum_{a=1}^d\,e^{-2\beta^a}{\cal
N}^{a}_{\,\,\,i}{\cal N}^{a}_{\,\,\,j}.\label{Iwasa}\ee The $\beta$'s are
often refered to as the scale factors although they more precisely
describe their logarithms. The ${\cal N}^a_{\,\,i}$'s measure the
strenght of the off-diagonal metric components and define how to pass
from the invariant $\{\omega^i\}$ co-frame to the Iwasawa co-frame
$\{\theta^a\}$ in which the metric is purely diagonal \be\theta^a = {\cal
N}^a_{\,\,j}
\omega^j.\ee In this basis, one has for the components of the $1$-form
\be A_j\equiv {\cal A}_{a}{\cal
N}^{a}_{\,\,\,j}.\label{AjAa}\ee The
changes of variables (\ref{Iwasa}) and (\ref{AjAa}) are continued to the
momenta as canonical point transformations in the standard way via \be
\pi^{ij}{\dot g}_{ij} +
\pi^{j}{\dot A}_{j} = \pi_a
\,{\dot\beta}^a + \sum_{a<j}{\cal P}^j_{\,\,a}\,{\dot{\cal
N}}^a_{\,\,j} + {\cal
E}^{a}\,{\dot {\cal A}}_{a}.\ee In this expression,
${\cal P}^j_{\,a}$ denotes the momentum conjugated to ${\cal N}^a_{\,j}$
and is defined for $a<j$, ${\cal E}^{a}$ denotes the momentum
conjugated to ${\cal A}_{a}$.  The Iwasawa components of
the electric and magnetic fields are given by
\be {\cal E}^{a}\equiv {\cal
N}^{a}_{\,\,\,\,j}\,\pi^{j}\quad,\quad {\cal F}_{a b}
\equiv F_{ij}\,{\cal N}^{i}_{\,\,\,\,a}{\cal
N}^{j}_{\,\,\,\,b}\ee where
${\cal N}^j_{\,\,a}$ denotes the element on line-$j$, column-$a$ of the
inverse matrix
${\cal N}^{-1}$. This matrix enters the definition the vectorial frame
$\{e_a\}$ dual to the co-frame $\{\theta^a\}$ by
\be e_{a} = {X}_j\,{\cal N}^j_{\,\,a}.\label{cbas}\ee While shifting to
the Iwasawa basis and co-basis, the structure constants of the group,
which also define the Lie brackets of the vectorial frame
$\{X_i\}$ dual to the invariant co-frame $\{\omega^i\}$
\be [ X_i, X_j] = - X_k\,C^k_{\,\,ij},\ee transform as the components of a
$(^1_2)$-tensor so that
\be [e_b, e_c] = -e_a\,C^{'a}_{\,\,\,\,bc},
\quad\mbox{with}\quad C^{'a}_{\,\,\,\,bc} = {\cal N}^{a}_{\,\,\,i}\, {\cal
N}^{j}_{\,\,\,b}\,{\cal N}^{k}_{\,\,\,c}\,C^i_{\,\,\,jk}.\label{struc}\ee

\subsection{Splitting of the Hamiltonian}

Following \cite{DHN}, we next split the Hamiltonian
$H$ into two parts: the first one, denoted by $H_0$,
is the kinetic term for the local scale factors $\beta^a$; the second
one, denoted by $V$, is the potential and contains all the other
contributions.  Thus
\be H = H_0 + V\ee and \be H_0 =
\frac{1}{4}\,G^{ab}\,\pi_a\,\pi_b.\ee The total potential
naturally splits into \be V = V_S + V_G + V^{el}_{(p)} +
V^{magn}_{(p)}\ee where
\be V_S = \frac{1}{2}\,\sum_{a<b}\,e^{-2(\beta^b -
\beta^a)}\,({\cal P}^j_{\,\,a}\,{\cal N}^b_{\,\,j})^2\label{pre}\ee is
quadratic in the ${\cal P}$'s  and as such related to the kinetic energy
of the off-diagonal metric components; one refers to it as to the
"symmetry" potential. $V_S$ vanishes in the case of pure diagonal
$g_{ij}$. Next comes the gravitational or curvature potential
\be V_G = - gR =
\frac{1}{2}\,e^{-2\sum_d
\beta^d}\sum_{a,b,c}( e^{2\beta^c}\,C^{\prime a}_{\,\,\,bc}\,C^{\prime
b}_{\,\,\,ac} +
\frac{1}{2}\,e^{-2\beta^a +2\beta^b + 2\beta^c}\,(C^{\prime
a}_{\,\,\,bc})^2\,)
\ee involving the structure constants in the Iwasawa basis defined by
(\ref{struc}). The last two terms in the potential correspond to the
electric and
magnetic energy: \beq V^{el}&=& \frac{1}{2}{\cal
E}^{a}\,{\cal E}_a =
\frac{1}{2}\,e^{-2e_{a}}\,({\cal
E}^{a})^2
\\ V^{magn} &=& \frac{1}{4}\,e^{-2\sum_{c=1}^d
\beta^c}\,{\cal F}_{a b}\,{\cal F}^{
a b}\nonumber
=\frac{1}{4}\,e^{-2 m_{a b}}\,({\cal
F}_{a b})^2
\eeq where, with the notations of \cite{DHN}),
\be
e_{a} =
\beta^{a}\quad,\quad m_{a b} =
\sum_{c\notin\{a b\}} \beta^c. \ee

\subsection{BKL Limit, billiard walls}

As the above formulae explicitly show, i) the
total potential exhibits the general form
\be V= V(\beta, {\cal N}, {\cal P}, {\cal E}, {\cal F}) = \sum_{A}\,c_A(
{\cal N}, {\cal P}, {\cal E}, {\cal F})\,exp (-2 w_A(\beta))\ee where
$w_A(\beta) = w_{Ab}\beta^b$ are linear forms in the scale factors and ii)
apart from the first term in the right-hand side of the curvature
potential (which will introduce subdominant walls), the prefactors $c_A$
are all given by the square of a real polynomial, implying $c_A\geq 0$. As
explained in \cite{DHN}, with the following gauge choice,
$\tilde N =
\rho^2 = -\beta_a\beta^a = -\rho^2 \gamma_a\gamma^a$, and in the BKL
limit corresponding to $\rho \to \infty$, the exponentials terms in the
potential
$\tilde N V$ become sharp walls and may be replaced by
\be
\lim_{\rho\to\infty}[c_A\,\rho^2\,e^{-2 \rho\,w_A(\gamma)}] =
\Theta_\infty (-2w_A(\gamma))\ee where $\Theta_\infty$ is defined
through
\be \Theta_\infty(x) = \left\{\begin{array}{ll}
  0\,\, \quad  x<0 \\
\infty \quad  x>0 \end{array}\right.\ee and has the property of being
invariant under multiplication by a positive factor. Accordingly, in this
limit, the positive $c_A$'s can be absorbed by the
$\Theta_\infty$ and the null ones can simply be dropped out; once this
has been done, the Hamiltonian no longer depends on the variables
$\chi$ \be \chi
\in \{{\cal N}, {\cal P}, {\cal E}, {\cal F}\}\ee that enter these
coefficients. This also means that the BKL Hamiltonian only contains the
scale factors
$\beta^a$ and their conjugate momenta $\pi_a$ hence, as can
be immediately infered from their equations of motion, all the
 $\chi$'s  must
asymptotically tend to constants. These constants are arbitrary
untill and unless the constraints enter the play.

Another way to see that, before the constraints are taken into
account,
the asymptotic values of the
$\chi$'s are arbitrary constants  is to consider the map \be
\phi: \chi_0\to
\chi^{\infty}= \phi(\chi_0)\ee which sends the initial values $\chi_0$ on
their asymptotic BKL values, for a given set of initial data $\{\beta_0,
\pi_0\}$, and to use the fact that, at least locally, this map must be
invertible. This statement is clearly true for an Hamiltonian which
does not at all depend on the $\chi$'s. It is reasonable to assume that this
property remains valid in the more difficult situation that we face here, where
the
$\chi$'s only asymptotically get out of the Hamiltonian.

In the
general inhomogeneous
 case, the constraints are essentially
conditions on the space gradients of the dynamical variables and, as
such, they introduce no limitation on the generic BKL Hamiltonian: that's
why generically the Hamiltonian contains all the walls. The same happens
of course when the constraints are absent. In the spatial homogeneity
context, however, they have to be taken into account since they
can influence the billiard's shape. Of course, the
prefactors controling the presence of the curvature walls crucially
depend on the homogeneity group under consideration: the more the group
"looks" abelian, the less curvature walls are present. The momentum
constraints further establish linear relations between the other wall
coefficients
$c_A$'s in which all together the structure constants, the $\chi$'s and
even the variables
$\{\beta, \pi\}$ are mixed up; so the
question arises whether, asymptotically, they are equivalent to the
condition that some of the
$c_A$'s vanish forcing the
corresponding walls to disappear. That is what we will systematically
investigate in the following.

\section{$d=3$ homogeneous
models}
\setcounter{equation}{0} \setcounter{theorem}{0}
\setcounter{lemma}{0}
\subsection{Iwasawa variables}
In spatial
dimension
$d=3$, using the simplified notations
\beq{\cal N}^1_{\,\,2} &=& n_1,\quad {\cal N}^1_{\,\,3} = n_2,\quad {\cal
N}^2_{\,\,3} = n_3\\ {\cal P}^2_{\,\,1} &=& p_1,\quad {\cal P}^3_{\,\,1} =
p_2, \quad{\cal P}^3_{\,\,2} = p_3,\eeq the prefactors $({\cal
P}^j_{\,a}{\cal N}^b_{\,j})^2$ of the possible symmetry walls
$e^{-2(\beta^b - \beta^a)}, b>a$, in (\ref{pre}), read
\beq
\mbox{for}\quad a=1, b=2 &:&\quad c_{12} =  (p_1 + n_3 p_2)^2
\label{coef1}\\
\mbox{for}\quad a=1, b=3 &:&\quad c_{13} =
(p_2)^2\label{coef2}\\ \mbox{for}\quad a=2, b=3 &:&\quad
c_{23} =  (p_3)^2.\label{coef3}\eeq The Iwasawa decomposition
(\ref{Iwas}) provides explicitly
 \beq g_{11} &=&
e^{-2\beta^1},\quad g_{12} = n_1 e^{-2\beta^1}, \quad g_{13} = n_2
e^{-2\beta^1}\label{g11}\\ g_{22} &=&n_1^2 e^{-2\beta^1} +
e^{-2\beta^2},\quad g_{23} = n_1 n_2 e^{-2\beta^1} + n_3
e^{-2\beta^2}\label{g22}\\ g_{33} &=& n_2^2
e^{-2\beta^1} + n_3^2 e^{-2\beta^2} +
e^{-2\beta^3}\label{g33}\eeq and its canonical extension reads \beq
2\pi^{11} &=& -(\pi_1 + 2 n_1 p_1 + 2 n_2 p_2) e^{2\beta^1} - (n_1^2
\pi_2 + 2 n_1^2 n_3 p_3 - 2 n_1 n_2 p_3)e^{2\beta^2} \nonumber \\ &-&
(n_2^2+ n_1^2 n_3^2 - 2 n_1 n_2 n_3) \pi_3 e^{2\beta^3}\\ 2\pi^{12}
&=& p_1 e^{2\beta^1} + (n_1 \pi_2 + 2 n_1 n_3 p_3 - n_2 p_3)
e^{2\beta^2}\nonumber\\ &+& (n_1 n_3^2 - n_2 n_3)\pi_3 e^{2\beta^3}
\\ 2\pi^{13} &=& p_2 e^{2\beta^1} - n_1 p_3 e^{2\beta^2} + (n_2 - n_1
n_3)\pi_3 e^{2\beta^3}\\ 2\pi^{22} &=& -(\pi_2 + 2n_3 p_3)
e^{2\beta^2} - n_3^2 \pi_3 e^{2\beta^3} \\ 2\pi^{23} &=& p_3
e^{2\beta^2} + n_3 \pi_3 e^{2\beta^3}\\ 2\pi^{33} &=& -\pi_3
e^{2\beta^3}.\eeq In order to easily translate the constraints in
terms of the Iwasawa variables, we also mention the following
usefull formulae \beq  2\pi^2_{\,1} &=& p_1\\ 2\pi^3_{\,1} &=& p_2\\
2\pi^3_{\,2} &=& n_1 p_2 + p_3\\ 2\pi^1_{\,2} &=& n_1 (\pi_2 -\pi_1) +
(e^{-2(\beta^2-\beta^1)}-n_1^2) p_1 + (n_3
e^{-2(\beta^2-\beta^1)} -
 n_1 n_2) p_2 \nonumber\\&+& (n_1 n_3 - n_2) p_3\\ 2\pi^1_{\,3} &=&
n_2(\pi_3-\pi_1) + n_1 n_3(\pi_2-\pi_3) + [n_3
e^{-2(\beta^2-\beta^1)} - n_1 n_2] p_1
\nonumber \\&+& [e^{-2(\beta^3-\beta^1)} + n_3^2 e^{-2(\beta^2-\beta^1)}-
n_2^2] p_2 \nonumber \\&+& [n_1 n_3^2 -n_2 n_3 - n_1
e^{-2(\beta^3-\beta^2)}] p_3
\\ 2\pi^2_{\,3} &=& -n_3 (\pi_2-\pi_3) + n_2 p_1 +
(e^{-2(\beta^3-\beta^2)}-n_3^2)p_3\eeq

\subsection{$d=3$ pure gravity billiards}
These spatially homogeneous models are known in the literature as the
class-A Bianchi models; they are classified according to their real,
unimodular, isometry Lie algebra \cite{LL,RS}.
\newline

\noindent 1. {\bf Bianchi-type I}: $C^k_{\,ij} = 0, \forall i,j,k$.

This is the abelian algebra. There is no spatial curvature and the
constraints are identically verified. Accordingly, the billiard
walls are only made of symmetry walls $\beta^i - \beta^j, i>j$,
among which the two dominant ones are $w_{32}=\beta^3-\beta^2$ and
$w_{21}=\beta^2-\beta^1$. This is the infinite volume non-diagonal
Kasner billiard. Its projection on the Poincar\'e disc is
represented by the shaded area in figure 1.
\newline

\noindent 2. {\bf Bianchi-type II}: $C^1_{\,23} = 1$.

This case is particularly simple because the constraints are easy to analyse.
Indeed, the momentum constraints read \be \pi^2_{\,1} = 0\quad\mbox{and}\quad
\pi^3_{\,1}=0\quad
\Longleftrightarrow\quad p_1 = 0\quad\mbox{and}\quad p_2=0.\ee That
means, referring to (\ref{coef1}), (\ref{coef2}) and (\ref{coef3}), that
$c_{12} =
0$ and $c_{13}=0$ and that they clearly eliminate the symmetry walls
$w_{21}=\beta^2 -
\beta^1$ and $w_{31}=\beta^3 -
\beta^1$; the last $p_3$ remains free so the
symmetry wall $w_{32} = \beta^3-\beta^2$ is present. Moreover, the only
non zero structure constants being
$C^{\prime 1}_{\,\,23} = 1$, one single curvature wall
survives which is
$2\beta^1$. Since two walls can never close the billiard, its volume is
also infinite. Its projection on the Poincar\'e disc is given by the shaded
area in figure 1.

\begin{figure}[h]
\centerline{\includegraphics[scale=0.6]{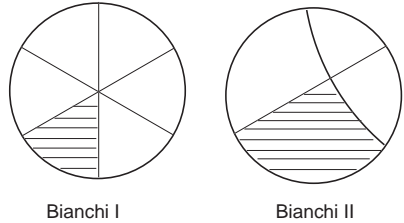}}
\caption{Bianchi I and II billiards are the shaded areas limited by curved
lines
or curvature walls and right lines or symmetry walls.}
\end{figure}

\noindent 3. {\bf Bianchi-type VI$_0$}: $ C^1_{\,23} = 1 = C^2_{\,13}$.

The momentum constraints read \be \pi^3_{\,2} = 0\quad,\quad
\pi^3_{\,1}=0\quad\mbox{and}\quad \pi^2_{\,1} + \pi^1_{\,2} =
0.\label{cos}\ee They are equivalent to \beq 0&=& p_2 \label{p1} \\ 0 &=&
p_3
\label{p2}\\ 0 &=& n_1 (\pi_2 -\pi_1) + (e^{-2(\beta^2-\beta^1)}-n_1^2 +2)
p_1 .\label{p3}\eeq The first two, according to (\ref{coef2}) and
(\ref{coef3}), clearly tell that the symmetry walls
$w_{32}=\beta^3-\beta^2$ and
$w_{31}=\beta^3-\beta^1$ are absent from the potential. With
(\ref{p2}) put in (\ref{coef1}), one sees that the coefficient of the
third symmetry wall
$w_{21}$ becomes $c_{12}=(p_1)^2$ and the question arises whether
the constraint (\ref{p3}) implies $p^\infty_1=0$? In order to
answer this question, we will exhibit an asymptotic solution for which
$p^\infty_1$ is a constant $\ne 0$. Hence, $p_1^\infty$
can be different from zero and the symmetry wall $w_{21} = \beta^2-\beta^1$
that
it multiplies is generically present.

Let us build such a solution. We know that after a finite number of
collisions \cite{DH} against the curvature walls, the ball will never meet them
again. At such a time, one can find a solution of Kasner's type with
ordered exponents. Let the diagonal Kasner
solution be\footnote{Here, the Kasner exponents will be denoted $q_i$ in
order to avoid confusion with $p_i$, the momenta conjugate
to
$n_i$}
\be ds^2=-dt^2+  \sum_{i=1}^{3}\,t^{2q_i}(dx^i)^2 \ee
this is an asymptotic solution of the  Bianchi VI type if $q_1
>0$ and $q_2>0$. On this diagonal asymptotic solution, we next perform a
linear transformation which generates another asymptotic solution
which is no longer diagonal: let
\be G= L^T \, G_{Kasner} \,L;\ee
$G$ is the new metric, $G_{Kasner}$ is the diagonal metric given
above. We select $L$ in the form of a Lorentz
boost\footnote{$L$ is not really restricted here to define a Lie algebra
automorphism, that is, the structure constants need not be
conserved because, at the limit considered, the curvature becomes
negligible and remains negligible after the transformation; however the
$L$-transformation here chosen conserves the Bianchi VI
structure constants and is a gauge transformation \cite{HC}.} in
the plane $(1,2)$ such as to produce but a single off-diagonal
element, namely $g_{12}$:

\be L= \left(  \begin{array}{ccc}
\cosh \alpha & \sinh\alpha&0\\
\sinh\alpha  & \cosh \alpha &0 \\
0&0&1\\
\end{array}  \right) .\ee Accordingly, the new metric elements are\beq
g_{11} &=& t^{2q_1}\cosh^2 \alpha + t^{2q_2}\sinh^2\alpha \\
g_{12} &=& (t^{2q_1}+ t^{2q_2})\sinh\alpha\cosh\alpha \eeq the
other stay unchanged. From these expressions and (\ref{g11}),
(\ref{g22}), (\ref{g33}), one immediately extracts the
$t$-dependence of $n_1$ and of the $\beta$'s which explicitly
gives \be n_1= \frac{g_{12}}{g_{11}} = \frac{(t^{2q_1}+t^{2q_2})\,
\cosh\alpha\sinh\alpha}{t^{2q_1}\cosh^2\alpha+t^{2q_2}\sinh^2\alpha}\ee
and \be e^{2(\beta^2-\beta^1)}=
\frac{(t^{2q_1}\cosh^2\alpha+t^{2q_2}
\sinh^2\alpha)^2}{t^{2q_1+2q_2}}.\ee In order to discover the time
behaviour of $p_1$, we make use of the equation of motion for
$n_1$ which, since $p_1$ just appears in the Hamiltonian through
the coefficient $c_{12}$, simply reads \be \dot{n}_1 = \frac{d
n_1}{d\tau}= \frac{d n_1}{dt}\frac{dt}{d\tau}=  p_1
e^{-2(\beta^2-\beta^1)},\ee remembering that \cite{DHN}
$dt=-\sqrt{g}\, d\tau$ and that the sum of the Kasner exponents is
$1$ such that $\sqrt{g}=t$. Hence, the momentum $p_1$ is given by
$p_1= -\dot{n_1}\, e^{2(\beta^2-\beta^1)}$; assuming $q_1 < q_2$,
one finds that, in the vicinity of $t=0$, it behaves as \be p_1
\simeq C + B \, t^{2(q_2-q_1)}+...\ee where $C= p_1^\infty$ is not
constrained to be zero. This result allows us to state that, in
general, the symmetry wall $w_{21}=\beta^2-\beta^1$ is present,
beside the two curvature walls $2\beta^1$ and $2\beta^2$; the
dominant walls being $w_{21}=\beta^2-\beta^1$ and $2\beta^1$, they
don't close the billiard.

Notice that, for $t\to 0$, the exponential in the constraint
(\ref{p3}) goes to zero like
\be e^{-2(\beta^2-\beta^1)}\simeq \cosh^{-4}\alpha \,t^{2(q_2 -q_1)},\ee
as announced by the BKL analyzis when the coefficient of the
wall $c_{12}\ne 0$ and also that the limit
$t\to 0$ of the product
$p_1\,exp -2(\beta^2-\beta^1)$ exists and is zero. Consequently,
(\ref{p3}) admits a clear limit when $t\to 0$.

\begin{figure}[h]
\centerline{\includegraphics[scale=0.6]{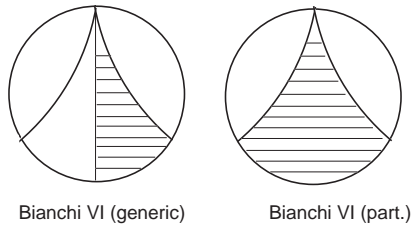}}
\caption{Bianchi VI Billiards}
\end{figure}

Had one taken advantage of the gauge freedom to assign the initial values
$n^0_1$
and
$p^0_1$ such as
$n^\infty_1=0$, then the constraint (\ref{p3}) would have imposed
the last symmetry wall to
be absent and the billiard, here limited by the two curvature
walls, to be open.

These examples show that the presence/absence of symmetry walls may depend
on initial conditions or on gauge conditions;  nevertheless, in the
cases mentionned hereabove, a billiard with an infinite volume remains
a billiard with an infinite volume.
\newline

\noindent 4. {\bf Bianchi-type VII}: $C^1_{\,23} = 1,\quad C^2_{\,13}=-1$.

This case is very similar to the preceding one. The only changes are
i) that the third constraint in (\ref{cos}) has to be replaced by
($\pi^2_{\,1} -
\pi^1_{\,2} = 0$) and ii) that its translation into the Iwasawa variables
in (\ref{p1}) now reads
\be 0 = n_1 (\pi_2
-\pi_1) + (e^{-2(\beta^2-\beta^1)}-n_1^2 -2) p_1.\ee Accordingly,
we can apply the same reasoning as before and conclude that
generically, $p_1^\infty \ne 0$ and that the symmetry wall
$w_{21}=\beta^2-\beta^1$ appears in the potential.
\newline

\noindent 5. {\bf Bianchi-type IX}: $C^1_{\,23} = 1,\quad C^2_{\,31} =
1,\quad C^3_{\,12} = 1$.

This case and the next one deserve a particular treatment because the structure
constants are such that all curvature walls, namely $2\beta^1$, $2\beta^2$,
$2\beta^3$, appear  and these three gravitational walls already form a
finite\footnote{This situation is very specific to the homogeneous models in
$D=4$; in higher spacetime dimensions, the number of curvature walls allowed by
the structure constants is not sufficient to produce a finite volume
billiard.} volume billiard.

Let us first discuss the generic case. The momentum
constraints take the form
\be
\pi^3_{\,2} -\pi^2_{\,3}= 0,\quad
\pi^3_{\,1}-\pi^1_{\,3} =0,\quad\pi^2_{\,1} - \pi^1_{\,2}
= 0;\ee and in terms of the Iwasawa variables, they become
\beq
 n_3 (\pi_2 &-&\pi_3)  - n_2 p_1 + n_1 p_2 \nonumber\\ &+&
[-e^{-2(\beta^3-\beta^2)}+n_3^2 +1]p_3 = 0 \label{b91}\eeq \beq
n_2(\pi_3-\pi_1) &+& n_1 n_3(\pi_2 -\pi_3)  + [n_3
e^{-2(\beta^2-\beta^1)} - n_1 n_2] p_1
\nonumber \\&+&[e^{-2(\beta^3-\beta^1)} + n_3^2 e^{-2(\beta^2-\beta^1)}-
n_2^2 - 1] p_2 \nonumber \\&+& [n_1 n_3^2 -n_2 n_3 - n_1
e^{-2(\beta^3-\beta^2)}] p_3 =0 \label{b92}\eeq  \beq  n_1 (\pi_2
&-&\pi_1) + [e^{-2(\beta^2-\beta^1)}-n_1^2- 2] p_1\nonumber \\&+& [n_3
e^{-2(\beta^2-\beta^1)}-
 n_1 n_2] p_2 \nonumber\\ &+& [n_1 n_3 - n_2] p_3=0.\label{b93}
\eeq Remember that we already know that the billiard has a finite volume, hence
the question is no longer to state between chaos or non chaos but rather to
define
more precisely the shape of the billiard. We cannot copy the reasoning made for
Bianchi VI since we have here to account for an infinite number of collisions.

In order to study the implications of the constraints (\ref{b91}) -
(\ref{b93}),
we shall rely on the heuristic estimates made in \cite{DHN}, where the
asymptotic
behaviour of the variables is analysed in the BKL limit. From that analysis, it
follows that, when $\rho\to \infty$: i) the $\pi_a$'s go to zero as powers of
$1/\rho$, ii) the $n_i$'s and the $p_i$'s tend to constants $n_i^\infty$ and
$p_i^\infty$ up to additive terms which also go to zero as powers of
$1/\rho$ and
iii) that the exponentials either vanish (if the walls are present) or
oscillate
between zero and
$\infty$ (if they are absent). What we infer from this information is that
if the
$p_i^\infty$ were not strictly zero then the constraint system hereabove would
have no BKL limit. This is obviously wrong because a constraint need be
obeyed all the time; so the limit must exist. Now we know that the $p_i$'s
go to
zero, can we say something more about the walls that they multiply? Compared to
the exponential growth of the wall, the power decrease of its prefactor is not
fast enough to prevent the symmetry wall to appear in the Hamiltonian.

Accordingly, the billiard's edge is formed by the leading symmetry walls
$w_{32}$ and $w_{21}$ and by the curvature wall $2\beta^1$. On the
Poincar\'e disc, it is one of the six small triangles included in the larger
one bordered by the curvature walls. Its Cartan matrix is that of the
hyperbolic
Kac-Moody algebra
$AE_3 = A_1^{\wedge\wedge}$ \be \left
(\begin{array}{ccc}2&-1&0\\-1&2&-2\\0&-2&2\end{array}\right )\ee already
relevant in the general inhomogeneous case. Its Dynkin diagram is given in
figure 3.

\begin{figure}[h]
\centerline{\includegraphics[scale=0.6]{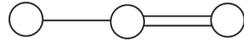}}
\caption{Dynkin diagrams of the $A_1^{\wedge\wedge}$ Kac-Moody algebra}
\end{figure}

Other interesting cases exist with less symmetry walls, which require specific
initial conditions.

\begin{figure}[h]
\centerline{\includegraphics[scale=0.6]{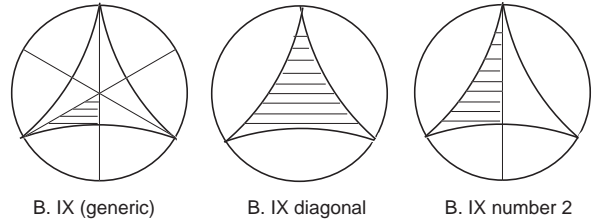}}
\caption{Bianchi IX Billiards}
\end{figure}

\begin{enumerate}
\item{No symmetry wall at all.} This situation is the one mentionned above;
it happens when the metric is assumed diagonal always, hence for
$n_1=0, n_2=0, n_3 = 0$; the solution of the momentum constraints is then
$p_1=0, p_2=0, p_3=0$. This assumption is consistent with the equations of
motion. The Cartan matrix of this billiard is given by
\be \left (\begin{array}{ccc}2&-2&-2\\-2&2&-2\\-2&-2&2\end{array}\right
)\ee and its corresponding Dynkin diagram is number 1 in figure 5; the
associated Kac-Moody algebra is hyperbolic, it has number 7 in the
enumeration provided in reference \cite{S}.

\item{One symmetry wall.} This happens when one chooses the initial data
such that
$n_1=0, n_2=0$ and $p_1=0$, $p_2=0$; then, according to the equations of
motion, these variables remains zero all the time. The billiard is closed
by two curvature walls
$2\beta^1$ and
$2\beta^2$ and the symmetry wall $w_{32}$. On the Poincar\'e disc, its
volume is half of that of the triangle made of curvature walls. The
Cartan matrix is given by
\be \left (\begin{array}{ccc}2&0&-2\\0&2&-2\\-2&-2&2\end{array}\right
)\ee and its Dynkin diagram is number 2 in figure 5. It characterizes the
third rank 3 Lorentzian Kac-Moody algebra in the classification given in
\cite{S}.
\end{enumerate}

\begin{figure}[h]
\centerline{\includegraphics[scale=0.6]{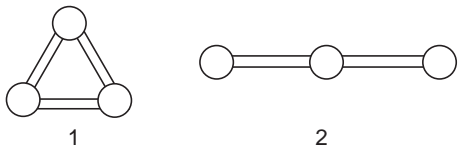}}
\caption{Dynkin diagrams of special algebras met in Bianchi
IX models}
\end{figure}

These last two Kac-Moody algebras are subalgebras of $A_1^{\wedge\wedge}$
\cite{FN}.

Remark: let us recall that the reflections on the faces of the billiard
generate a Coxeter group which is identified with the Weyl group of the
associated Kac-Moody algebra. The larger the set of walls, the
larger the number of generators. Since in the non
generic cases the set of walls is a subset of the one in the generic case,
the associate Coxeter group is a subgroup of the generic one.
\newline

\noindent 6. {\bf Bianchi-type VIII}: $C^1_{\,23} = 1,\quad C^2_{\,31} =
1,\quad C^3_{\,12} = -1$.

The analysis of this case follows closely the
previous one. The sign change in the structure constants
modifies the constraints as follows
\be
\pi^3_{\,2} +\pi^2_{\,3}= 0,\quad
\pi^3_{\,1}+\pi^1_{\,3} =0,\quad\pi^2_{\,1} - \pi^1_{\,2}
= 0\ee and induces some sign changes in their Iwasawa counterparts.
The conclusions are those of the Bianchi IX model: the generic billiard is the
one of the algebra $A_1^{\wedge\wedge}$.

\subsection{$d=3$ Einstein-Maxwell billiards}

The momentum constraints (\ref{momentum}) of the Einstein-Maxwell
homogeneous models generally (except for the abelian Bianchi I) mix
gravitational and one-form variables; in comparison with the pure
gravity case, i) no constraint remains which clearly forces the
prefactor of a symmetry wall to be zero and ii) additional terms of the type
$\pi_{(1)} F^{(1)}$ enter in the constraints system.

Accordingly, generically, besides the curvature walls allowed by
the structure constants, one expects all symmetry, electric and
magnetic walls to be present. The dominant ones are the symmetry
walls $w_{21}= \beta^2-\beta^1, w_{32}= \beta^3-\beta^2$ and the
electric wall $e_1=\beta^1$ which replaces the curvature wall of
the pure gravity case. They close the billiard whose Cartan matrix
is \be \left
(\begin{array}{ccc}2&-4&0\\-1&2&-1\\0&-1&2\end{array}\right ).\ee
The Dynkin diagram is dispayed in figure 6; the corresponding
Kac-Moody algebra is the hyperbolic $A_2^{(2)\wedge}$ algebra
which is the Lorentzian extension of the twisted affine algebra
$A_2^{(2)}$ also encountered in \cite{HJ}.

\begin{figure}[h]
\centerline{\includegraphics[scale=0.6]{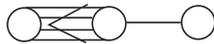}}
\caption{Dynkin diagram of the $A_2^{(2)\wedge}$ algebra}
\end{figure}

\section{$d=4$ homogeneous models}
\setcounter{equation}{0} \setcounter{theorem}{0}
\setcounter{lemma}{0}
Our billiard analysis of the four dimensional spatially homogeneous
cosmological models has been carried out along the same lines as for
$d=3$: results relative to pure gravity models and those relative to the
Einstein-Maxwell homogeneous system will be given separately.

\subsection{Extension of the notations and Iwasawa variables}

In spatial dimension $d=4$, we introduce the matrix variables \be
{\mathcal N} =
\left (\begin{array}{cccc}
1&n_{12}&n_{13}&n_{14}\\0&1&n_{23}&n_{24}\\
0&0&1&n_{34}\\0&0&0&1\end{array}\right).\ee The Iwasawa decomposition
of the metric extends beyond the 3-dimensional formulae listed in
(\ref{g11}), (\ref{g22}), (\ref{g33}), through the additional
components  \beq g_{14} &=& n_{14}\,e^{-2\beta^1}\\ g_{24} &=&
n_{12}\,n_{14}\,e^{-2\beta^1} + n_{24}\,e^{-2\beta^2}\\ g_{34} &=&
n_{13}\,n_{14}\,e^{-2\beta^1} + n_{23}\,n_{24}\,e^{-2\beta^2}+
n_{34}\,e^{-2\beta^3}\\ g_{44}&=& n_{14}^2\,e^{-2\beta^1} +
n_{24}^2\,e^{-2\beta^2} + n_{34}^2\,e^{-2\beta^3} + e^{-2\beta^4}.
\eeq
The momentum conjugate to $\beta^a$ is written as $\pi_a$, as before; the
momentum conjugate to
${\cal N}^a_{\,j} = n_{aj}$ is denoted
${\cal P}^j_{\,a} = p^{ja}$ and is only defined for $j>a$.
The prefactors $c_{ab}=({\cal
P}^j_{\,a}{\cal N}^b_{\,j})^2$ of the symmetry walls $e^{-2(\beta^b
-\beta^a)}, b>a,$
in the Hamiltonian
 are explicitly given by \beq c_{12} &=& (p^{21} +
p^{31}\,n_{23} + p^{41}\,n_{24})^2
\\c_{13}&=& (p^{31} + p^{41}\,n_{34})^2\\c_{14}&=&
(p^{41})^2\\c_{23}&=& (p^{32} + p^{42}\,n_{34})^2\\c_{24}&=& (p^{42})^2\\
c_{34}&=& (p^{43})^2.\eeq
Once the change of dynamical variables has been continued in a
canonical point transformation, the momentum contraints
still express linear relations among the $\pi^i_{\,j}$'s which
translate into linear relations on the momenta
$\pi_a, p^{ia}$; their
coefficients are polynomials in the $n_{ai}$'s times
exponentials of the type
$e^{-2(\beta_b -\beta_a)}$, with $b>a$. The explicit form of the
constraints depends on the model considered.

\subsection{$d=4$ Pure gravity}

We label the various 4 dimensional
spatially homogeneous models according to the classification of the 4
dimensional real unimodular Lie algebras given by M. MacCallum \cite{McC}:
they are

\begin{enumerate}
\item{\bf Class $U1[1,1,1]$}: $C^1_{\,\,14}=\lambda,\,
C^2_{\,\,24}=\mu,\,C^3_{\,\,34}=\nu, \mbox{with}\,\,\lambda+\mu+\nu=0$.
One can still set $\lambda=1$ except in the abelian case where
$\lambda=\mu=\nu=0$.

\item{\bf Class $U1[Z,\bar Z,1]$}: $C^1_{\,\,14}=-\frac{\mu}{2},\,
C^2_{\,\,14}=1, C^1_{\,\,24}=-1,\, C^2_{\,\,24}= -\frac{\mu}{2},\\
C^3_{\,\,34}=\mu.$ $\mu=0$ is a special case.

\item{\bf Class $U1[2,1]$}: $C^1_{\,\,14}=-\frac{\mu}{2},\,
C^2_{\,\,14}=1, \, C^2_{\,\,24}= -\frac{\mu}{2},\,C^3_{\,\,34}=
\mu $ and $\mu$ is $0$ or
$1$.
\item{\bf Class $U1[3]$}:
$C^2_{\,\,14}=1, \, C^3_{\,\,24}= 1$.
\item{\bf Class $U3I0$}: $C^4_{\,\,23}=1,\,
C^2_{\,\,31}=1, \, C^3_{\,\,12}= -1.$
\item{\bf Class $U3I2$}: $C^4_{\,\,23}=-1,\,
C^2_{\,\,31}=1, \, C^3_{\,\,12}= 1.$
\item{\bf Class $U3S1$} or $s\ell(2)\oplus u(1)$: $C^1_{\,\,23}=1,\,
C^2_{\,\,31}=1, \, C^3_{\,\,12}= -1.$
\item{\bf Class $U3S3$} or $su(2)\oplus u(1)$: $C^1_{\,\,23}=1,\,
C^2_{\,\,31}=1, \, C^3_{\,\,12}= 1.$

\end{enumerate}

Let us mention that for all of the four dimensional algebras, except of
course the abelian one, the structure constant $C^{\prime
1}_{\,\,\,34}\ne 0$ and consequently that the curvature wall
$2\beta^1+\beta^2$ is always present.

Our analysis leads to the conclusion that, from the billiard point of
view, the previous models can be collected into two sets: the first one
has an open billiard, the second one has a finite volume billiard
whose Cartan matrix is that of the hyperbolic Kac-Moody algebra
$A_2^{\wedge\wedge}$ exactly as in the general inhomogeneous situation.

The first set contains the abelian algebra and $U1[1,1,1]_{\mu\ne
0,-1}$, $U1[2,1]$ and $U1[Z,\bar Z,1]_{\mu\ne 0}$; all the others belong to
the second set. Because explicit developments soon become lenghty and
since the reasonings always rest on similar arguments, we have chosen
not to review systematically all cases as for $d=3$ but rather to
illustrate the
results on examples taken in both sets:
\newline

\noindent{1. As a representative of the first set, we take
$U1[Z,\bar Z,1]_{\mu\ne 0}$}.

The momentum constraints read \be \pi^4_{\,1}=\pi^4_{\,2} =
\mu\,\pi^4_{\,3} = -\frac{\mu}{2}(\pi^1_{\,1} + \pi^2_{\,2}
-2\,\pi^3_{\,3}\,) + \pi^1_{\,2} - \pi^2_{\,1} = 0;\label{co1}
\ee in terms of the Iwasawa variables, they become
\be p^{41} = p^{42} = \mu\,p^{43} = 0\label{co2}\ee and \beq
&-&\frac{\mu}{2} (2\pi_3 - \pi_1
-\pi_2) + n_{12} (\pi_2 -\pi_1) +
\nonumber\\ &+&(e^{-2(\beta^2-\beta^1)}-n_{12}^2-1) \, p^{21}
+(n_{23}e^{-2(\beta^2-\beta^1)}- n_{12}n_{13}+\frac{3 n_{13}\mu}{2}  )
p^{31}
\nonumber \\ &+& (n_{12}n_{23}-
n_{13}+\frac{3 n_{23}\mu}{2}
)\,p^{32}=0.\label{co3}\eeq
The first three constraints (\ref{co2}) clearly indicate that the
symmetry walls $\beta^4-\beta^3$,  $\beta^4-\beta^2$,
$\beta^4-\beta^1$ are absent from the potential. What information can one
extract from the fourth constraint (\ref{co3})? In order to answer this
question, we will proceed as for the Bianch VI model: we will
exhibit an asymptotic solution with prefactors
$c_{23} = (p^{32}_{\infty})^2$ and  $c_{12}= (p^{21}_{\infty}+
n_{23}^{\infty}\,p^{31}_{\infty})^2$ different from zero. We here again face
a situation in which the billiard motion is made of a finite number of
collisions
against the curvature walls. After the last bounce, the motion can be
described by
a Kasner type solution
\be ds^2=-dt^2+  \sum_{i=1}^{4}\,t^{2q_i}\,(dx^i)^2\ee
which is a suitable asymptotic solution
at the conditions  $q_1
>0$, $q_2>0$ and $q_3>0$ (no condition on $q_4$ is required because there
is no curvature wall involving $\beta^4$). We next perform a linear
transformation  \be G= L^T G_{Kasner}L\ee in order to
get another asymptotic solution which is non-diagonal. Here, for
simplicity, $L$ can be
taken in the form that produces the off-diagonal
elements we are intersted in,
\be L= \left( \begin{array}{cccc}
l_1 &  l_2&l_3&0\\
m_1 & m_2 & m_3&0  \\
r_1&r_2&r_3&0\\
0&0&0&1\\
\end{array}  \right).\ee
Then  using the equations of motion for $n_{12}$, $n_{13}$  and
$n_{23}$ we obtain
\be p^{21}+n_{23}\,p^{31} = \dot{n_{12}}\,
e^{2(\beta^2-\beta^1)} \label{p211}\ee and
\be p^{32} = \dot{n_{23}}\,  e^{2(\beta^3-\beta^2)}.\label{p212}\ee
By direct application of the formulae given in reference
\cite{DHN}, section (4.2), assuming the following order $q_1<q_2<q_3$
for the Kasner exponents, we find that for
$t\to 0$, (\ref{p211}) and (\ref{p212}) tend to constants that depend on
the parametrization of $L$ but that will generically be different from zero.

Consequently, our conclusion is that, in the
generic case, the dominant walls of the billiard are the symmetry walls
$w_{32}=\beta^3-\beta^2$,
$w_{21}=\beta^2-\beta^1$ and  the curvature wall $2\beta^1+\beta^2$; it is
indeed an open billiard.

Notice that in all cases of the first set, the non vanishing  structure
constants are all of the form $C^{\prime a}_{\hspace{.15cm}4b}$  with
$a,b=1,2,3$. It is easy to check that this forbids all
curvature wall containing $\beta^4$ and therefore that the remaining curvature
walls cannot be expected to close the billiard.
\newline

\noindent{2. As a first representative of the second set, we take
$U1[Z,\bar Z,1]_{\mu = 0}$}

The constraints are given by (\ref{co1}), (\ref{co2}) and
(\ref{co3}) for $\mu=0$ and one immediately sees that, compared to
the preceding case, one constraint drops out leaving the symmetry
wall $w_{43}=\beta^4-\beta^3$ in place. To conclude the analysis
of the last constraint, we can also provide a solution with non
vanishing prefactors for the symmetry walls. Since the structure
constants do not play an important r\^ole in the explicit
construction hereabove, the solution obtained with $\mu\ne 0$ can
also be used for $\mu=0$ if one assigns particular initial
condition such that $p^{43}=0$ and $n_{34}=0$. Again, we end up
here with the following list of dominant walls:
$w_{43}=\beta^4-\beta^3$, $w_{32}=\beta^3-\beta^2$,
$w_{21}=\beta^2-\beta^1$ and $2\beta^1+\beta^2$. The Cartan matrix
is that of the algebra $A_2^{\wedge\wedge}$ as previously
announced.

\begin{figure}[h]
\centerline{\includegraphics[scale=0.5]{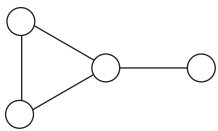}}
\caption{Dynkin diagram of the $A^{\wedge\wedge}_2$ algebra}
\end{figure}

\noindent{3. Another interesting example of the second set is provided by
$U3S3$}

Its homogeneity group is the
direct product
$SU(2)\times U(1)$ so that this model appears as the four dimensional
trivial extension of the Bianchi IX model: since the structure constants
are the same as in the 3-dimensional case, the momentum constraints do not
change, when expressed in terms of the metric components and their momenta
\be
\pi^1_3-
\pi^3_1=0,\quad
\pi^1_2-\pi^2_1=0,\quad
\pi^2_3- \pi^3_1=0;\ee their number does not change either but, in
terms of Iwasawa variables, they involve much more terms than their $d=3$
counterpart given in (\ref{b91}), (\ref{b92}) and (\ref{b93})
\beq &-&n_{13}(\pi_1-\pi_3) + n_{12} n_{23}(\pi_2  -\pi_3)
-(n_{12}n_{13}-n_{23}e^{-2(\beta^2-\beta^1)})p^{21}\nonumber \\ &-&
(n_{13}^2-n_{23}^2e^{-2(\beta^2-\beta^1)}-e^{-2(\beta^3-\beta^1)}-1)
p^{31}\nonumber\\
&+&(n_{12}n_{23}^2-n_{13}n_{23}-n_{12}e^{-2(\beta^3-\beta^2)})p^{32}\nonumber
\\ &-&(n_{13}n_{14}-n_{23}e^{-2(\beta^2-\beta^1)}n_{24}-n_{34}
e^{-2(\beta^3-\beta^1)})p^{41}\nonumber \\
&+&(n_{12}n_{23}n_{24}-n_{14}n_{23}-n_{12}n_{34}e^{-2(\beta^3-\beta^2)})
p^{42}\nonumber\\
&-&(n_{12}n_{23}n_{34}-n_{12}n_{24}-n_{13}n_{34}+n_{14})p^{43}=0 \eeq \beq
&-& n_{12} (\pi_1- \pi_2)  - (n_{12}^2-e^{-2(\beta^2-\beta^1)}-1 )p^{21}
\nonumber
\\&-& (  n_{12}  n_{13}- n_{23}e^{-2(\beta^2-\beta^1)} )  p^{31}
+ (n_{12}  n_{23}- n_{13} ) p^{32} \nonumber\\ &-& ( n_{12} n_{14}-
n_{24} e^{-2(\beta^2-\beta^1)})p^{41}+ (  n_{12} n_{24}- n_{14}  )
p^{42}=0 \eeq \beq &-& n_{23} (\pi_2- \pi_3)+ n_{13}
p^{21}- n_{12}  p^{31}-( n_{23}^ 2 - e^{-2(\beta^3-\beta^2)}+1
)p^{32}\nonumber \\ &-& (  n_{23}  n_{24} - n_{34}e^{-2(\beta^3-\beta^2)})
p^{42}+ ( n_{23} n_{34}- n_{24} )  p^{43}=0.\eeq
These constraints are of course i) linear in the momenta, ii) the
only exponentials which enter these expressions are build of
$w_{32}=\beta^3-\beta^2$,
$w_{31}=\beta^3-\beta^1$ and
$w_{21}=\beta^2-\beta^1$ and iii) as before, their coefficients are
exactly given by the square root of the corresponding $c_A$'s in the
potential, namely
$\sqrt{c_{12}},
\sqrt{c_{13}}$ and
$\sqrt{c_{23}}$. Accordingly, the question arises whether these equations are
equivalent to $c_{12}=0$, $c_{23}=0$ and $c_{13}=0$. If, asymptotically, the
exponentials go to zero and  can be dropped out of the constraints, the
remaining equations can be solved for $p^{21}, p^{31}, p^{32}$ in
terms of the other variables among which figure now, not only the
$\pi_i-\pi_j$ already present in the 3-dimensional case which
asymptotically go to
zero, but also the asymptotic values of the
$p^{4i}$'s which remain unconstrained. It follows that none of the
solutions of the above system is generically forced to vanish so that all the
symmetry walls are expected to be present. The absence of a wall can only
happen
in non generic situations with very peculiar initial conditions.

We  expect this result to become the rule
in higher dimensions for trivial extensions like $SU(2)\times
SU(2)\times...\times U(1)
\times...\times U(1)$; the
billiard will then become that of the Kac-Moody algebra
$A_n^{\wedge\wedge}$ relevant in the general Einstein theory.

We can nevertheless provide a particular solution with initial data obeying
$n_{12}=n_{13}=n_{14}=0$ and $p^{21}=p^{31} = p^{41}=0$. These values are
conserved in the time evolution. In this case, the leading walls are the
symmetry walls $w_{43}, w_{32}$ and the curvature ones $2\beta^2+\beta^3,
2\beta^1+\beta^2$. The Cartan matrix is
\be \left( \begin{array}{cccc}
2 & -1&-1&0\\
-1 & 2 & -1&-1 \\
-1&-1&2&-1\\
0&-1&-1&2\\
\end{array}  \right).\ee The billiard is characterized by the rank
4 Lorentzian Kac-Moody algebra which bears number 2 in the classification
given in
\cite{S}; its Dynkin diagram is drawn in figure 8.

\begin{figure}[h]
\centerline{\includegraphics[scale=0.5]{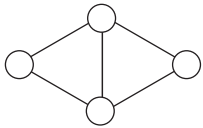}}
\caption{Dynkin diagram of algebra number 2 in the subset of rank 4}
\end{figure}

The $D=5$ results are summarized in the following table:
\newline

\begin{center}

\begin{tabular}{|p{3cm}|c|}
\hline  $U1 [ 1,1,1 ] _{\mu = \nu = \lambda =0}$ $U1[1,1,1]_{\mu
\neq 0,  -1}$
$U1[Z,\bar{Z},1]_{\mu \neq 0}$& non chaotic \\
\hline  $U1[1,1,1]_{\mu = -1}$ $U1[Z,\bar{Z},1]_{\mu = 0}$
$U1[2,1]$\newline $U1[3]$
\newline
$U3I0$  \newline $U3I2$ \newline $U3S3$ \newline $U3S1$ & chaotic  \\
\hline
\end{tabular}
\end{center}
\centerline{Chaos or non chaos for $D=5$ models}

\subsection{$d=4$ Einstein-Maxwell models}

As in the three dimensional case, because of the presence of the
electromagnetic field in the expression of the momentum  constraint
(\ref{momentum}), no symmetry wall can be eliminated. Moreover, the electric
walls always prevail over the curvature ones if any.
The billiard is in all cases
characterized by the following set of dominant walls
$w_{43}=\beta^4-\beta^3$, $w_{32}=\beta^3-\beta^2$,
$w_{21}=\beta^2-\beta^1$ and
$e_1=\beta^1$. Its Cartan matrix is that of the hyperbolic
Kac-Moody algebra
$G^{\wedge\wedge}_2$.

\begin{figure}[h]
\centerline{\includegraphics[scale=0.6]{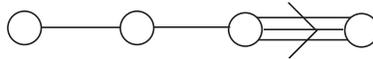}}
\caption{Dynkin diagram of the $G^{\wedge\wedge}_2$ algebra}
\end{figure}

In the general analysis of the billiards attached to coupled
gravity + p-forms systems, one could assume $2p<d$ without loss of
generality, because the complete set of walls is invariant under
electric-magnetic duality. This invariance may however not remain
in some spatially homogeneous cases due to the vanishing of some
structure constants which lead to incomplete sets of walls so that
other, a priori unexpected Kac-Moody algebras, might appear. An
illustrative and interesting example is given in $D=5$ by the
Einstein + $2$-form system governed by the abelian algebra. Here,
the dominant walls are  i) the symmetry walls $\beta^4-\beta^3$,
$\beta^3-\beta^2$ and $\beta^2-\beta^1$ and ii) the electric wall
$\beta^1+\beta^2$. This billiard brings in the new hyperbolic
Kac-Moody algebra carrying number 20 in \cite{S}, whose Dynkin
diagram is given in figure 10 hereafter.

\begin{figure}[h]
\centerline{\includegraphics[scale=0.6]{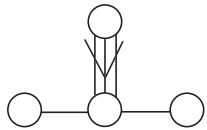}}
\caption{Dynkin diagram of the  algebra relevant for the Einstein
+ $2$-form system}
\end{figure}

\section{Conclusions}
\setcounter{equation}{0} \setcounter{theorem}{0}
\setcounter{lemma}{0}
In this paper, we  have analyzed the Einstein and Einstein-Maxwell
billiards for  all the  spatially homogeneous cosmological models
in 3 and 4 dimensions. In the billiard picture, we confirm that in
spacetime dimensions $5\leq D\leq10$, diagonal models are not rich enough
to produce the never ending oscillatory behaviour of the generic solution of
Einstein's equations and that chaos is restored when off-diagonal metric
elements
are kept. Chaotic models are characterized by a finite volume billiard
which can
be identified with the fundamental Weyl chamber of an hyperbolic Kac-Moody
algebra: in the most generic chaotic situation, the algebra coincides with
the one
already relevant in the inhomogeneous case; this remains true after the
addition of an generic homogeneous electromagnetic field. Other algebras
can also
appear for  special initial data or gauge choices: in fact, these are all the
simply-laced known hyperbolic Kac-Moody algebras of ranks 3 and 4, except
the one
which has number 3 among those of rank 4 listed in \cite{S}. The billiard of a
non chaotic model is even not a simplex.

\section*{Acknowledgements}
We are most grateful to Marc Henneaux for many informative
discussions and for essential hints. We also thank S. Hervik for
his careful reading of our manuscript, his comments and
corrections. This work is supported in part by the ``Actions de
Recherche Concert{\'e}es" of the ``Direction de la Recherche
Scientifique - Communaut{\'e} Fran{\c c}aise de Belgique", by a
``P\^ole d'Attraction Interuniversitaire" (Belgium), by
IISN-Belgium (convention 4.4505.86) and by the European Commission
RTN programme HPRN-CT-00131.

\end{document}